\documentclass[12pt]{article}
\usepackage{graphicx,footnpag}
\newcommand{\pt}{$p_{\rm T}$}
\newcommand{\pl}{$p_{\rm L}$}

\newcommand{\npart}{$N_{\rm part}$}
\newcommand{\ncoll}{$N_{\rm coll}$}
\newcommand{\dnde}{$dN_{\rm ch}/d\eta$}
\newcommand{\dndy}{$dN_{\rm ch}/dy$}
\newcommand{\nch}{$N_{\rm ch}$}
\newcommand{\dndemax}{$dN/d\eta \mid_{\rm max}$}
\newcommand{\dndymax}{$dN/dy \mid_{\rm y=y_{cm}}$}

\newcommand{\pp}{${\rm pp}(\bar{\rm p})$}
\begin{document}

\begin{center}
\Large
{\bf Charged particle pseudorapidity distributions in nucleus-nucleus 
collisions from SPS to LHC\\}
\medskip
\large
{Francesco Prino, INFN sezione di Torino}
\end{center}

\begin{abstract}
Pseudorapidity distributions of charged particles produced in heavy ion 
collisions are a powerful tool to characterize the global properties of the
created system.
They have been measured in a wide range of energies at the GSI, AGS, SPS 
and RHIC accelerators and will be one of the first measurements at
ALICE at the LHC.
The various analysis techniques developed by SPS and RHIC experiments
are reviewed and the presently available results
from SPS and RHIC are presented, focusing in particular on the scaling
of charged particle yield with centrality and with center-of-mass energy.
Finally the perspectives for the multiplicity measurement in
the ALICE experiment at the LHC are discussed.
\end{abstract}

\section{Introduction}

Heavy-ion collisions are the experimental tool used to study nuclear 
matter under extreme temperature and density conditions.
The total number of particles produced in these collisions (multiplicity) is a 
global variable that is essential for their characterization, 
because it quantifies to which extent the incoming beam energy is released to 
produce new particles.
Particle multiplicity contains information about the entropy of the 
system and the gluon density in the first stages of the collision evolution.
Furthermore, it is related to the impact parameter, i.e. to the centrality 
of the collision: more particles are produced in central (small impact 
parameter, many elementary nucleon-nucleon collisions) than in peripheral 
reactions.

Since experimental detection methods are usually sensitive to ionizing 
(charged) particles, it is useful to introduce the charged multiplicity 
(\nch) of the collision defined as the total number of charged particles 
produced in the interaction.

The measurement of charged multiplicity as a function of the energy 
($\sqrt{s}$) and the centrality of the collision
may help constrain different models of particle production, and estimate the 
relative importance of soft versus hard processes 
in the particle production mechanisms at different energies.
Hard parton-parton scatterings with large momentum transfer occur on a short 
time scale and are governed by perturbative QCD; they are expected to
scale like the number of elementary nucleon-nucleon collisions (\ncoll). 
The bulk of particle production occurs via soft processes (with
low momentum transfer and consequently longer time-scales) which are described
with phenomenological models and predicted to scale with the number of 
participant nucleons (\npart)~\cite{Bialas}.

\begin{figure}[tb]
\centering
\resizebox{\textwidth}{!}{
\includegraphics{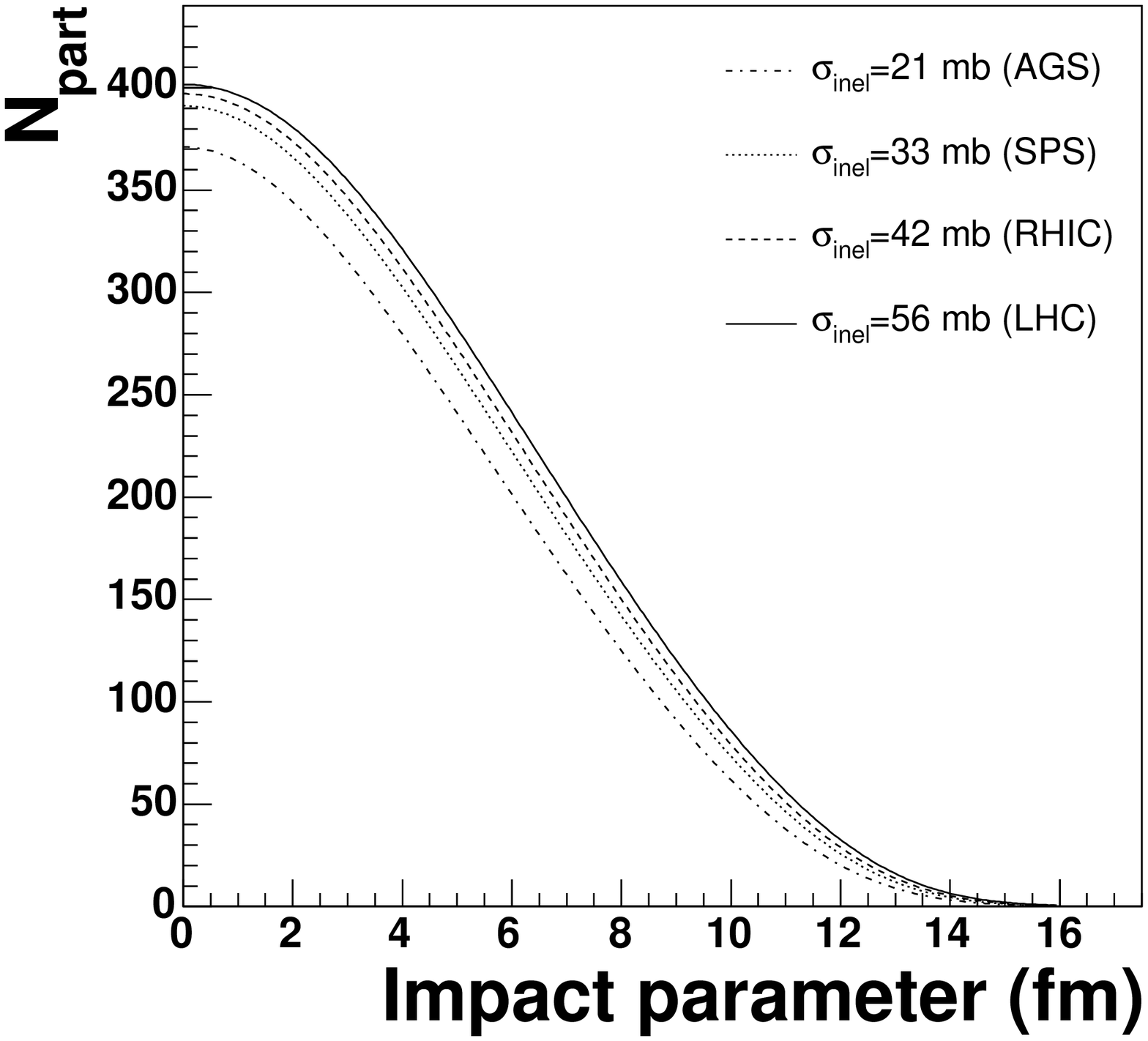}
\includegraphics{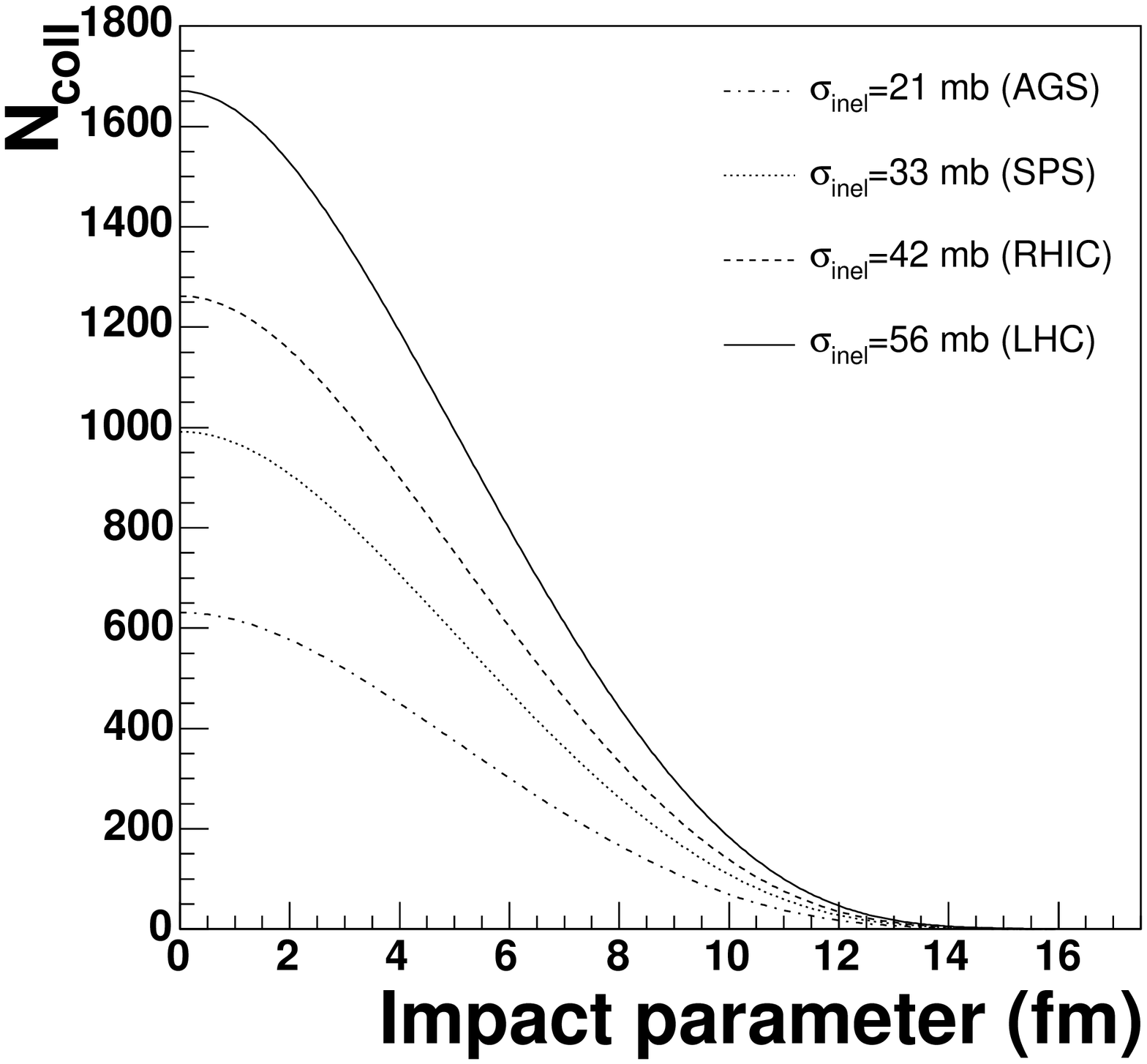}
\includegraphics{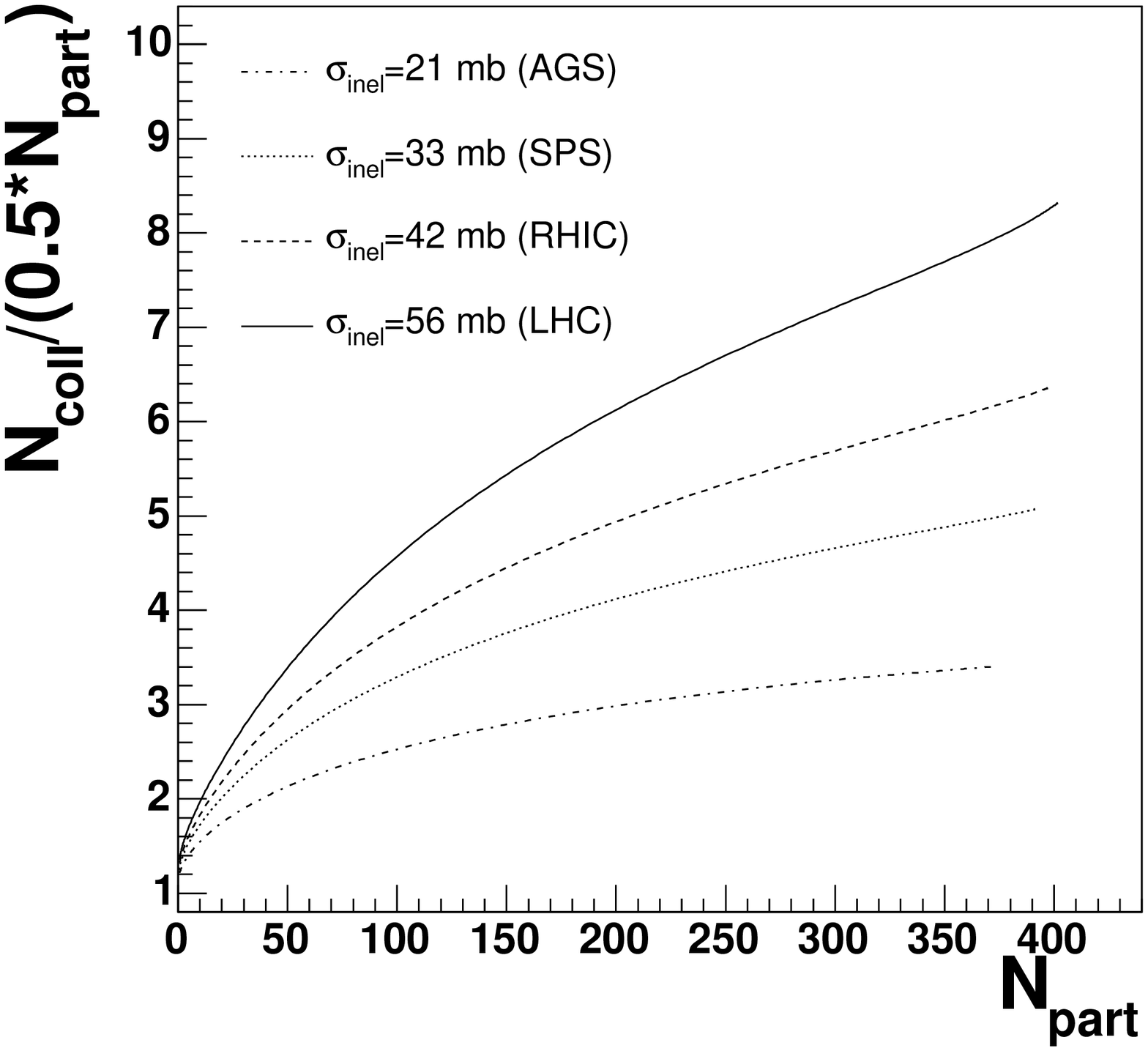}}
\caption{\npart\ (left) and \ncoll\ (middle) as a function of impact parameter
$b$ for different values of nucleon-nucleon inelastic cross-section.
In the right panel, the number of collisions per participant pair as a
function of \npart is shown.}
\label{fig:glau}
\end{figure}

The values of \npart\ and \ncoll\ as a function of the collision impact 
parameter are usually evaluated by means of Glauber model 
calculations~\cite{Glauber}.
An example of such calculations for Pb-Pb collisions and different values of
the nucleon-nucleon inelastic cross-sections is shown in fig.~\ref{fig:glau}.
It can be seen (left panel) that \npart\ does not change significantly from
AGS to LHC energies, meaning that the system volume is almost independent
of energy.
On the contrary \ncoll\ (middle panel), being proportional to the 
nucleon-nucleon inelastic cross-section, increases dramatically 
with increasing energy,
leading to the expectation that the higher the collision energy, 
the more important the contribution from hard processes.

From the right plot in fig.~\ref{fig:glau} it can be seen that the number of
collisions per participant pair (which governs the balance between
hard and soft processes), increases both with the centrality (\npart) and 
the energy of the collision.

A deeper insight into particle production mechanisms can be obtained from
particle momenta distributions.
Particle momenta are usually decomposed in the  
component transverse to the beam axis (\pt) and in the one along the 
beam axis (\pl).
In this way, all the information about the velocity of the particle-emitting
source is contained in the longitudinal momentum, while the transverse momentum
is free from kinematic effects and is governed only by the internal
characteristics of the system which emits the particles.

To study the longitudinal expansion it is convenient to use the rapidity
$y=\frac{1}{2}\ln\frac{E+p_L}{E-p_L}$
and pseudorapidity 
$\eta=-\ln \left[ \tan\left(\vartheta/2\right) \right]$
variables, the latter being an approximation of the rapidity
for large momentum particles.
Pseudorapidity is more easily accessed experimentally because it requires 
to measure only one kinematic quantity for each particle, i.e. the angle 
$\vartheta$ of relative to the beam axis.

In the following, pseudorapidity distribution of unidentified charged particles
will be discussed.
Since this measurement does not require momentum measurement and particle
identification capabilities, it is a typical first-day observable in 
heavy-ion experiments. 
Nevertheless, the following interesting physical issues can be addressed
by studying these distributions:
\begin{enumerate}
\item{The mid-rapidity region which is populated by 
particles with \pt$>$\pl\ (\pt=\pl\ corresponds to $\eta=\pm 0.88$)
is especially sensitive to the details of the hadroproduction mechanisms. 
The pseudorapidity density of charged particles at mid-rapidity (\dndymax\ 
or \dndemax) is commonly used to characterize the multiplicity of particles 
produced 
in the reaction because it allows a comparison between different 
experiments, being independent of the detailed phase-space acceptance.
Furthermore, under the assumption of a boost-invariant central plateau, it
can be used to estimate the energy density 
reached in the collision using the Bjorken formula~\cite{Bjorken}:
\begin{equation}
\varepsilon_{BJ}=
\frac{<m_T>}{\mathcal{A} c \tau_o} \left(\frac{dN}{dy}\right) _{y=y_{cm}}
\end{equation}

}
\item{The width of the pseudorapidity distributions contains information about
the longitudinal expansion of the system and the degree of stopping 
(transparency) which characterizes the reaction.
In the case of two nuclei that completely stop each other giving rise to a
baryon-rich fireball, a single isotropic source emitting at rest should be
observed and the \dnde\ distributions  should present a typical 
$1/\cosh^2 \eta$ shape, with a FWHM of 1.72 pseudorapidity units.
}
\item{The forward and backward regions where \pl$>>$\pt\ (close to beam 
rapidities) allow to investigate
effects connected with projectile and target fragmentation,
with particular attention to the issue of 
``limiting fragmentation''~\cite{Benecke}.}
\end{enumerate}

\section{Experimental issues}

Different analysis techniques have been developed by various experiments 
to extract the \dnde\ distributions of charged particles. 
In the following, a brief review of the methods used by SPS and RHIC 
experiments is given.

A first method, typically used with detectors with high segmentation and 
binary readout (only hit/no hit information for each channel),
consists in counting fired channels (hits).
Analyses based on this method have been performed by
NA50~\cite{NA50a} at SPS and PHOBOS~\cite{PHOBOSb} at RHIC.
Generally, one hit on a detector is not necessarily equal to one crossing
particle because of multiple occupancy (two or more particles hitting the same
sensitive element) and clustering effects which cause the signal of one
particle to extend to more than one detector channel.
Multiple occupancy can be reduced by increasing the detector segmentation
so as to keep the channel occupancy below a confidence level.
Clustering effects can occur both due to physical processes (such as
inclined tracks, charge sharing between contiguous sensitive elements) and to
electronic effects (noise, cross-talk between electronic channels...).
Hence, in experimental analyses based on detectors with binary readout, 
clusters (i.e. groups of contiguous channels
firing together) are counted and a correction to account for
multiple occupancy in a cluster is applied.
As an example, NA50 reported~\cite{NA50a} that a Monte Carlo simulation, 
which contains a complete description of their experimental setup and
only clustering mechanisms of physical origin, 
is not able to reproduce the cluster-size distributions observed
on their silicon microstrip detector.
Probabilities for clustering mechanisms were therefore included in the 
simulations so as to fit the measured cluster-size distributions and to
extract the number of incident particles.

If the detector readout stores the signal amplitudes, thus providing a 
measurement of energy deposition in each channel, the number of crossing 
particles can be calculated from the signal recorded in each channel 
divided by the average energy loss of a typical particle.
Particular attention has to be used when dealing with thin detectors (such 
as silicon strip detectors), where the energy deposition follows a Landau 
distribution and a correction for particles depositing much more energy than 
the most probable value has to be included in the analysis.
This method has been used by NA57~\cite{NA57} at SPS and PHOBOS~\cite{PHOBOSb} 
and BRAHMS~\cite{BRAHMSa,BRAHMSb}
at RHIC.

A more refined method can be used when two planes of highly segmented 
detectors are present. 
This method was first introduced by PHOBOS experiment~\cite{PHOBOSa} and 
consists in
associating hits on two detector planes and building a so-called tracklet. 
The association is normally done by considering straight line tracks departing 
from the primary vertex which is therefore required to be known.
A similar method has also been used by PHENIX~\cite{PHENIX} experiment.
Compared to the previous single-plane methods, tracklets allow a more powerful
background rejection but require more precision in detector alignment and
knowledge of the primary vertex.
 
Finally, a method based on full track reconstruction has been used by 
experiments with large tracking detectors, such as NA49~\cite{NA49a} and 
STAR~\cite{STARmult}.
This is the most powerful method, which is normally accompanied by
momentum measurement and particle identification, thus allowing
to measure rapidity distributions of the various particle species.

A correction for secondary production in material placed upstream of the
detector plane has to be applied, especially in the case of analyses based
on single plane measurement.
This correction is usually extracted from Monte Carlo simulations
of the experimental apparatus.
In the case of NA50, for example, the secondary/primary ratio is reported 
to be of the order of 1.2-1.8~\cite{NA50a}, mainly due to the Pb target 
thickness.
The analyses based on tracking require a Monte Carlo correction to take
into account tracking efficiency and, if a magnetic field is present, 
low \pt\ particle cut-off.

\section{Experimental results from SPS and RHIC}

In the following sections some of the presently available
experimental results from SPS and RHIC accelerators are discussed, focusing
on the centrality dependence at a given energy and on the 
energy dependence observed for the most central collisions.
First the pseudorapidity density of charged particles at mid-rapidity are 
presented.
Then, measurements of the width of the \dnde\ distributions are reviewed.
Subsequently, the main features observed in the fragmentation regions, 
with particular attention to the limiting fragmentation behaviour are 
described.
Finally, the total yield of charged particles is discussed.

\subsection{Particle yield at mid-rapidity}

\subsubsection{Scaling with centrality}

The scaling of pseudorapidity density of charged particles
at mid-rapidity as a function of centrality is an important test
for models of particle production in heavy ion reactions because it allows
to quantify the relative importance of soft 
($\propto$\npart) and hard ($\propto$\ncoll) processes.

SPS experiments typically performed \npart$^\alpha$ fits to the measured
\dndemax\ in different centrality bins.
Values of $\alpha$ ranging in the range from 1.00 (NA50~\cite{NA50b}) to 1.08
(WA98~\cite{WA98}) have been found.
This is usually understood as an indication that hard scatterings do not
play an appreciable role at such energies.
It has to be stressed that the value of $\alpha$ depends on the model used
to calculate \npart, which is not a direct experimental observable.
For example, NA50 quoted $\alpha=1.00$ when \npart\ is extracted from a 
Glauber calculation in the optical approximation and 
$\alpha=1.08$ when a Monte Carlo generator (VENUS~\cite{VENUS}) is used.

A convenient variable to be used in such studies is the particle yield per
participant pair at mid-rapidity, defined as \dndemax/(\npart/2).
At SPS energies this quantity is found to be rather flat
as a function of centrality, reflecting the approximate scaling with \npart\
of the multiplicity at mid-rapidity.

\begin{figure}[bt]
\centering
\resizebox{.5\textwidth}{!}{
\includegraphics*[]{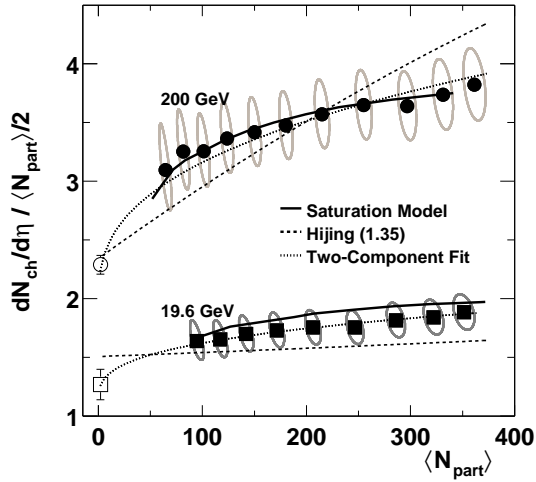}}
\caption{Mid-rapidity pseudorapidity density per participant pair as a 
function of centrality as measured by PHOBOS experiment at 
$\sqrt{s}$=19.6 and 200 GeV (taken from~\cite{PHOBOS2004})}
\label{centscal-max}
\end{figure}

At RHIC energies ($\sqrt{s}=20-200$ GeV) a clear increase is observed
in the \dndemax\ per participant pair with increasing centrality, 
as it can be seen in fig.~\ref{centscal-max}.
The normalized yield at mid-rapidity increases by $\approx$ 25\% from 
mid-peripheral to central collisions.
Early theoretical explanations attributed this increase to the contribution of 
hard processes in particle production, which grows with increasing centrality.
However, this explanation is challenged by the observed fact that the ratio of
the measured \dndemax\ per participant pair at the two energies is almost 
independent of centrality, while the contribution from hard processes should
lead to a stronger centrality dependence with increasing 
collision energy, as it is illustrated by the HIJING~\cite{HIJING} 
prediction shown in fig.~\ref{centscal-max}.
This consideration is also supported by the results of fits to the data points
in fig.~\ref{centscal-max} with a simple two-component 
parametrization~\cite{CGC1}:
\begin{equation}
\frac{dN_{ch}}{d\eta}=n_{pp}\left( (1-x)\frac{<N_{part}>}{2}+x<N_{coll}>
\right)
\end{equation}
The parameter $x$, representing the fraction of hard process, is found to
be consistent at both energies with a single value 
$x=0.13\pm0.01\pm0.05$~\cite{PHOBOS2004}.
The RHIC data about particle yields at mid-rapidity are described rather 
well by models based on parton saturation~\cite{CGC1,CGC2}, 
indicating that high density QCD effects probably play an important role in 
determining the global event features at RHIC energies.
However, as pointed out in~\cite{STARwp}, different models to calculate \npart\
(which is not a direct experimental observable and affects both axes of 
fig.~\ref{centscal-max}) would lead to different slopes for the centrality 
dependence of the \dndemax\ per participant pair, thus weakening the
relevance of parton saturation and Color Glass Condensate in the initial state 
of RHIC collisions.
By the way, differences in \npart\ calculation may explain also why the 
yield per participant pair is found to be almost flat 
as a function
of centrality at $\sqrt{s}$=17.2 GeV by SPS experiments and
increasing with centrality by PHOBOS at the very similar energy of
 $\sqrt{s}$=19.6 GeV.

\subsubsection{Scaling with energy}

Important information can be extracted analyzing the pseudorapidity 
density of charged
particles at mid-rapidity for central events as a function of the 
center-of-mass energy of the collision.
When comparing the \dndemax\ values between fixed-target and collider
experiments, it should be taken into account that \dnde\ (contrarily to \dndy)
distributions are not boost invariant and therefore the value at the peak
differs if measured in the laboratory or in the center-of-mass frame.
So, data measured in the laboratory frame should be converted into the Lorentz
invariant \dndy\ and then back to the \dnde\ in the center-of-mass using the 
formula:
\begin{equation}
\frac{dN_{ch}}{d{\bf p}_T d\eta}= 
\sqrt{1-\frac{m^2}{m^2_T \mathrm{cosh}^2 y}}~\frac{dN_{ch}}{d{\bf p}_T dy}
\end{equation}

\begin{figure}[tb]
\centering
\resizebox{.6\textwidth}{!}{
\includegraphics*[]{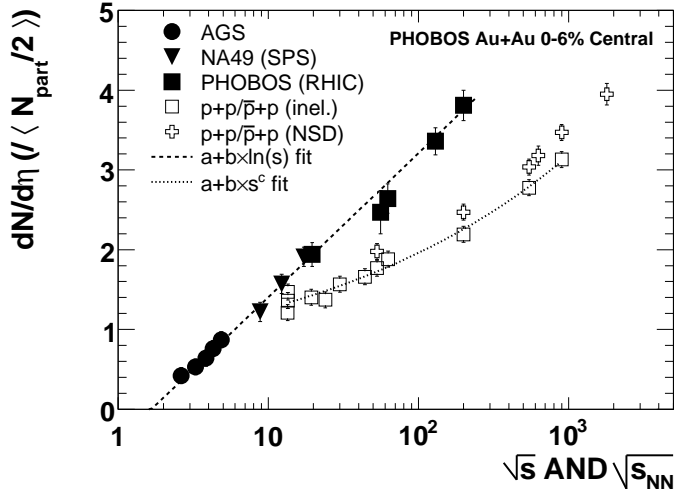}}
\caption{Energy dependence of the 
pseudorapidity density of charged particles per participant pair 
(in the center-of-mass frame) for the most central ion-ion collisions from
AGS to RHIC (taken from~\cite{PHOBOS05}). \pp\ data are superimposed. }
\label{enscal-max}
\end{figure}

The scaling of \dndemax\ per participant pair as a function of $\sqrt{s}$ 
for central heavy-ion collisions from AGS to RHIC energies is shown in 
fig~\ref{enscal-max} together with the same quantity as measured in 
proton-proton collisions.
It is immediately seen that nucleus-nucleus collisions can not be 
explained as an incoherent superposition of nucleon-nucleon interactions.
The multiplicity per participant pair in nucleus-nucleus collisions 
is in agreement with the one measured in \pp\ reactions only at AGS energies. 
In the SPS energy range a departure from
the proton-(anti)proton trend is observed and more particles per participant 
pair are produced at mid-rapidity in nuclear collisions with respect to 
proton collisions at the same energy.
The measured multiplicities appear to fall on a smooth curve from
AGS to top RHIC energies and this result contrasts with some theoretical 
predictions made before RHIC startup, which were suggesting strong energy 
dependences accompanying the hadron to QGP phase transition.

Starting from the measured values of \dndemax\ an estimation of the
energy density attained in central Au-Au collisions at top RHIC energy can be
performed using the Bjorken formula:
\begin{equation}
\varepsilon_{BJ}=
\frac{<m_T>}{\mathcal{A} c \tau_o} \left(\frac{dN}{dy}\right) _{y=y_{cm}}
\approx \frac{0.6 {\rm GeV/c^2}}{145 {\rm fm^2}\times c\times \tau_o}
\times \left(700\times \frac{3}{2}\times1.1\right)
\end{equation}
where the factor 3/2 accounts for neutral particles and the factor 1.1
for the difference between \dnde\ and \dndy.
Depending on the value assumed for $\tau_o$, i.e. on the instant when the 
energy density is evaluated, different values of $\varepsilon$ are obtained.
At the formation time $\tau_f=\hbar/<m_T>\approx 0.35$ fm/c~\cite{PHENIXwp} 
the energy density would be $\approx$15 GeV/fm$^3$.
At the instant $\tau=0.6-1$ fm/c, which is the time 
estimated for the system to reach local thermal equilibrium, the energy 
density would be $\approx$5-9 GeV/fm$^3$.
It is clear that, also for the more conservative choice for $\tau_o$,
the estimated energy density for Au-Au collisions at $\sqrt{s}$=200 GeV is 
well above the one predicted by lattice QCD calculations
for the phase transition to QGP.

\subsection{Width of the distribution}

\subsubsection{Scaling with centrality}

A decrease of the width of the \dnde\ distributions (expressed as gaussian width or FWHM) 
with increasing collision centrality has been observed by several experiments at 
various energies, among which E802~\cite{Abb92} at the AGS, 
NA50~\cite{NA50b} at two SPS energies and PHOBOS~\cite{PHOBOS03b} at RHIC.
This narrowing can be associated with the higher degree of stopping 
reached in the interaction, and it is mostly due to the decreasing contribution of protons 
from target and projectile fragmentation. 
In fact, emulsion experiments, which report the distribution of shower particles 
$(\beta > 0.7)$ excluding therefore slow protons from the target fragmentation, 
usually find a weaker dependence of $\sigma_{\eta}$ on centrality (see e.g.~\cite{Ada94}).

\subsubsection{Scaling with energy}

\begin{figure}[tb ]
\centering
\resizebox{.55\textwidth}{!}{
\includegraphics*[]{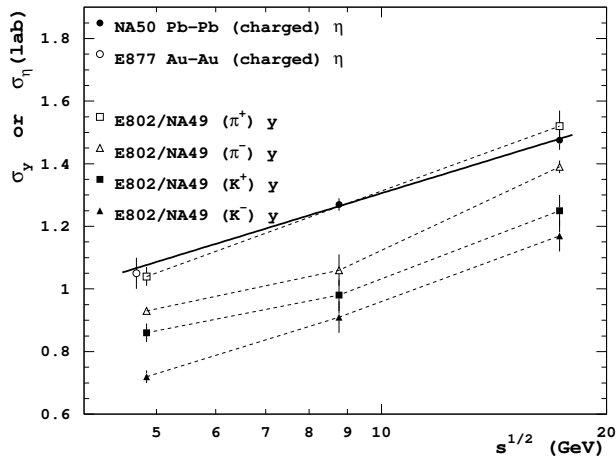}}
\caption{Energy dependence of the width of gaussian fits to pseudorapidity
and rapidity distributions at AGS and SPS energies (taken from~\cite{NA50b}).}
\label{fig:widthvsen}
\end{figure}

The width of the gaussian fits performed to \dnde\ (\dndy) distributions
in central collisions at SPS and AGS energies as a function of
$\sqrt{s}$ is shown in fig.~\ref{fig:widthvsen}.
It can be seen that the width increases with increasing $\sqrt{s}$
reflecting the fact that the available phase space in rapidity
increases with the center-of-mass energy,
following a simple logarithmic scaling law 
($\sigma_{\eta} = a + b \times \ln\sqrt{s}$) independent of system size.
.The same scaling is observed also for the widths of the rapidity 
distributions for identified produced hadrons with 
$\sigma_{\eta} \approx \sigma_y(\pi^+)$ and
$\sigma_y(\pi^+) > \sigma_y(\pi^-) > \sigma_y(K^+) > \sigma_y(K^-) $.

Finally, it is interesting to note~\cite{Sen99} that at 158 GeV/nucleon 
the width of the rapidity distributions is about twice as large 
as the one expected from a single thermal source located at mid-rapidity.

\subsection{Fragmentation regions}

Taking advantage of its large pseudorapidity coverage,
PHOBOS experiment
studied particle production in the fragmentation regions
of the colliding nuclei, so as to investigate the limiting fragmentation 
hypothesis~\cite{Benecke}.
For this analysis, it is convenient to consider the
particle distributions in the rest frame of one of the two colliding nuclei by
introducing the variables $y^\prime=y-y_{beam}$ and $\eta^\prime=\eta-y_{beam}$.
The limiting fragmentation ansatz states that at high enough collision 
energy, both $d^2N/dy^\prime dp_T$ and the mix of particle species reach a 
limiting value and become energy independent in a region around $y^\prime=0$.
In particular, this effect implies also a limiting value 
for $dN/d\eta^\prime$ which is expected to be energy independent in a 
region around $\eta^\prime=0$.

\begin{figure}[bt ]
\centering
\resizebox{.8\textwidth}{!}{
\includegraphics*[]{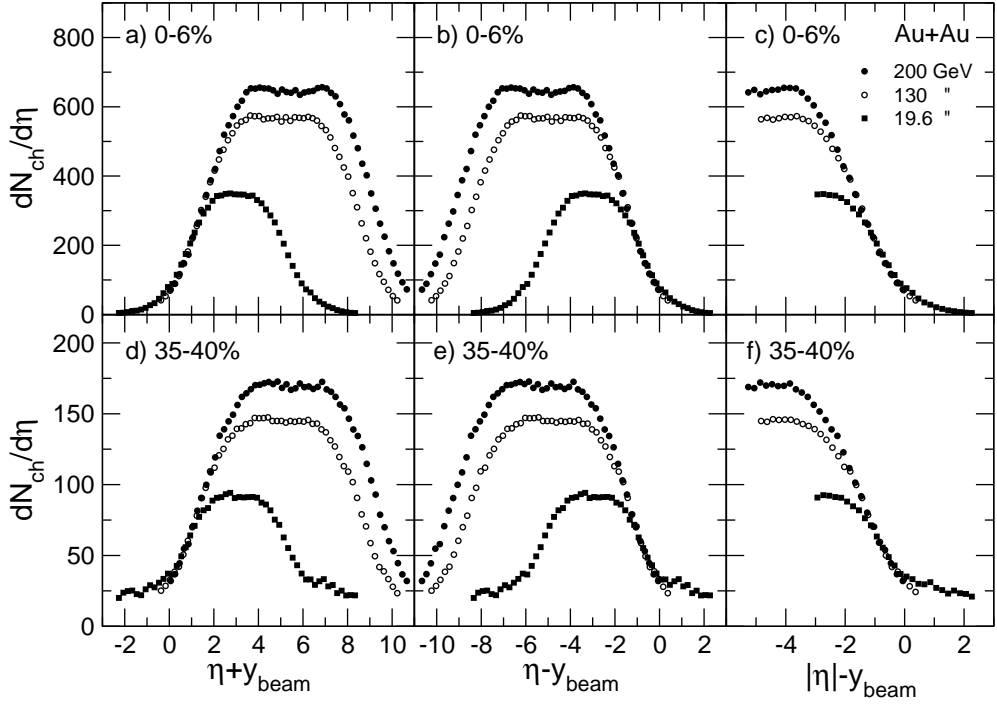}}
\caption{Pseudorapidity distributions of charged particles in 
Au-Au collisions at $\sqrt{s}$=19.6, 130 and 200 GeV and two centrality 
ranges in the rest frames of the two colliding nuclei 
(taken from~\cite{PHOBOSwp}).}
\label{fig:limfrag}
\end{figure}

The results obtained by PHOBOS~\cite{PHOBOS03b} in Au-Au collisions
at three different energies are shown in fig.~\ref{fig:limfrag}.
The distributions follow a common limiting curve 
independent of collision energy over a wide $\eta^\prime$ range.
Furthermore, the extent of the ``limiting fragmentation region'' grows
significantly with increasing energy and at $\sqrt{s}$=200 GeV it extends 
more than two units away from the beam rapidity.
This result is in contrast to the boost-invariance scenario~\cite{Bjorken}
which predicts a broad plateau at mid-rapidity growing in extent with 
increasing beam energy.
The measured pseudorapidity distributions appear to be dominated by two 
``fragmentation'' regions, whose extent increases with collision energy.
It can also be seen in fig.~\ref{fig:limfrag} that the ``limiting curve'' 
is different between central (top panels) and peripheral (bottom panels)
collisions.

\subsection{Integrated particle yield}

The total number of charged particle is calculated by extrapolating
the measured \dnde\ distributions over the full solid angle.
In the case of PHOBOS experiment, which covers the pseudorapidity
region $|\eta|<5.4$, the extrapolation is quite small even at the 
highest RHIC energy.
On the contrary, in the case of NA50, only $\approx$50\% of the particles
are in the acceptance of the multiplicity detector.

\subsubsection{Scaling with centrality}

\begin{figure}[bt]
\centering
\resizebox{.5\textwidth}{!}{
\includegraphics*[]{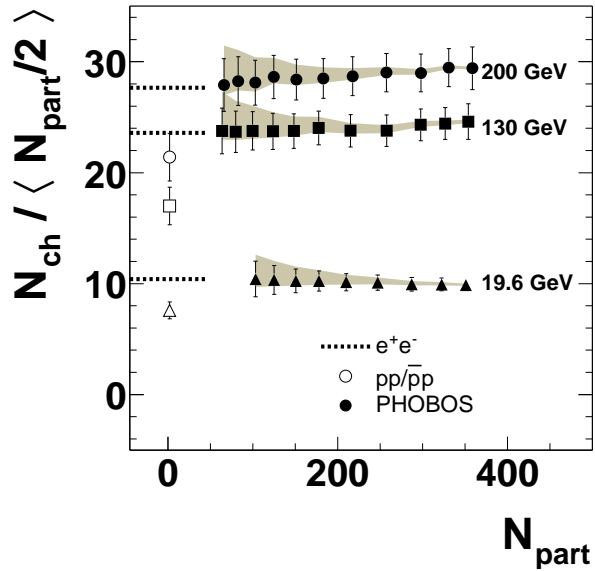}}
\caption{Centrality dependence of the total charged multiplicity 
per participant pair as measured by PHOBOS experiment at 
$\sqrt{s}$=19.6, 130 and 200 GeV (taken from~\cite{PHOBOS03}).}
\label{fig:centscal-tot}
\end{figure}

The total charged particle yield per participant pair
obtained by PHOBOS by integrating the measured
\dnde\ distributions is shown in fig.~\ref{fig:centscal-tot}
for different centrality bins at three different RHIC energies.
Contrarily to the particle yield at mid-rapidity, the
total charged-particle multiplicity results to be proportional to the
number of participant nucleons at all three energies from
$\sqrt{s}$=19.6 to 200 GeV.
This \npart\ scaling comes from a compensation between the
narrowing of the \dnde\ distributions and the
more than linear increase of \dndemax\ with increasing centrality.
Furthermore, it can be seen that the charged multiplicity per participant 
pair in Au-Au collisions at RHIC agrees with the one measured in 
$e^+e^-$ (and not in \pp) collisions at the same energy.

\subsubsection{Scaling with energy}

\begin{figure}[tb]
\centering
\resizebox{.6\textwidth}{!}{
\includegraphics*[]{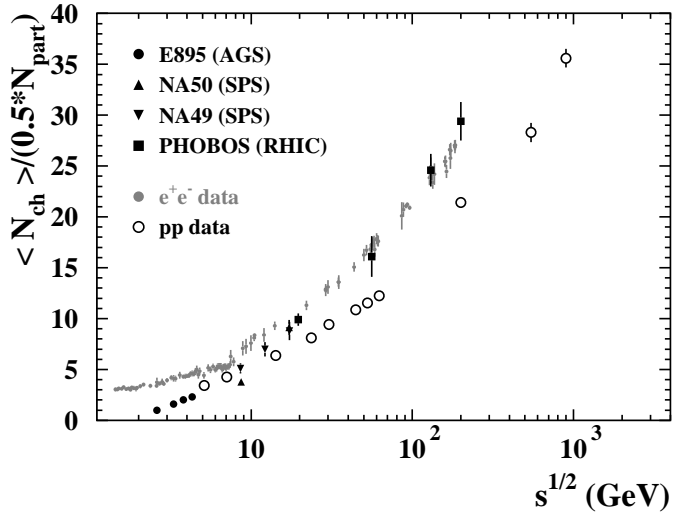}}
\caption{Energy dependence of the total charged multiplicity per participant 
pair for the most central ion-ion collisions from
AGS to RHIC. \pp\ and $e^+e^-$ data are superimposed.}
\label{fig:enscal-tot}
\end{figure}

The total charged multiplicity per participant pair as measured in 
central collision at
AGS, SPS and RHIC experiments is shown in fig.~\ref{fig:enscal-tot} together
with the same quantity from \pp\ and $e^+e^-$ collisions.

The fact that \pp\ multiplicities lie about 30\% below the $e^+e^-$ data 
at the same energy can possibly be explained by the 
``leading particle effect'': the outgoing proton 
carries away a substantial amount of the beam energy, 
thus reducing the energy effectively available for particle 
production~\cite{Basile}.
It is observed that the multiplicities measured in \pp\ and
$e^+e^-$ reactions become consistent over a broad energy range if the
effective energy  ($\sqrt{s_{eff}}=\sqrt{s}/2$) is introduced for
\pp\ collisions to account for the leading particle effect.

Heavy ion multiplicities do not follow the $e^+e^-$ trend over the 
whole $\sqrt{s}$ range. 
Instead, they lie below the \pp\ data at AGS energies, cross
through the \pp\ curve around $\sqrt{s}\sim 10$ GeV and then gradually join 
the $e^+e^-$ trend above the SPS energies. 
At RHIC energies, the multiplicity per participant pair in heavy-ion 
collisions agrees with the one observed in $e^+e^-$ reactions 
at the same $\sqrt{s}$,
suggesting a substantially reduced leading particle effect in central 
collisions of heavy nuclei at high energy~\cite{PHOBOS03}.
The ``suppression'' of A-A multiplicity with respect to $e^+e^-$
at low energies is possibly explained by the larger number of baryons 
produced at such energies which tend to suppress the overall multiplicity,
due to the fact that the baryon chemical potential reduces the 
entropy~\cite{PHOBOSwp}.
These considerations suggest that the total multiplicity per participant 
pair might be a universal function of the available energy, irrespectively of
the colliding system.

\section{Perspectives for ALICE at the LHC}

The predictions for the charged multiplicity at mid-rapidity in a central Pb-Pb 
collision at the LHC ($\sqrt{s}=5.5$ TeV) before RHIC startup varied 
between 2000 and 10000 particles per unit of pseudorapidity.
At that time, therefore, ALICE~\cite{ALICEPROP} detectors were designed for 
optimal performance up to \dndemax$\approx$ 5000 and reliable performance
for \dndemax\ up to 8000.

The wealth of multiplicity measurements which meanwhile became available
from SPS and RHIC showed that the increase of multiplicity with increasing
$\sqrt{s}$ was less dramatic than expected.
Extrapolations of the presently available \dndemax\ values measured 
in central collisions at AGS, SPS and RHIC (see fig.~\ref{enscal-max}) 
would give for LHC energy values in the range between 1100 and 2000 particles 
per rapidity unit depending on the assumption on the $\sqrt{s}$ dependence 
of \dndemax.
However, given the large energy gap between top RHIC and LHC energies,
a dramatic change in the balance between particle production mechanisms 
is predicted to occur: multiplicities higher than the ones obtained from
a simple extrapolation of RHIC data could be observed at the LHC due to a 
large contribution from mini-jets and hard scatterings.

Various detectors of the ALICE setup can be used to measure the charged 
particle multiplicity and to reconstruct the \dnde\ distributions over a wide 
$\eta$ range.
Different detecting and analysis techniques will therefore be used.

In the central rapidity region the \dnde\ distributions can be efficiently 
reconstructed using the two innermost layers of the Inner Tracking System 
(ITS)~\cite{ALICEITS} 
which consist of two cylindrical layers of Silicon Pixel Detectors 
(SPD) with a length along the beam axis of 28.2 cm and radii of 4 cm 
(layer 1) and 7.2 cm (layer 2) respectively.
These detectors will operate close to the beam pipe, in a region where
the track density could be as high as 80 tracks per cm$^2$.
The pixel size is 50$\mu$m in the r$\varphi$ direction and 425 $\mu$m along 
the beam axis, resulting in $\approx$ 9.8 million channels which provide a 
binary (hit/no hit) information.
Such a high segmentation allow to keep the occupancy below 1.5\% for layer 1 
and below 0.4\% for layer 2 also in the case of 8000 charged particles
in the central unit of $\eta$.
The SPDs are the detectors of the ALICE barrel which
provide the wider $\eta$ coverage in the region around mid-rapidity 
($|\eta|<2$ for layer 1, $|\eta|<1.4$ for layer 2).

Two different methods to estimate charged-particle multiplicity 
from SPD are considered.
The first one consists in counting the number of ``clusters'' on each of the 
two SPD layers, the second in counting the number of ``tracklets'' obtained 
by associating clusters in the two layers.
The performances of these methods have been evaluated by means of Monte Carlo
simulations based with a detailed description of the detector geometry.
The reconstructed values of \dndemax\ by counting clusters on layer 1
are found to reproduce the generated values for \dndemax$\leq$5000, while  
for higher multiplicities cluster merging effects lead to an 
underestimation of the number of generated particles.
The \dnde\ reconstruction based on counting clusters on layer 2 
suffers from secondary production in the inner layer and consequently 
gives an overestimation of the generated multiplicity.
The ``tracklet'' method has the advantage of allowing for 
an efficient background rejection (noise, secondary particles) by defining 
appropriate $\Delta \eta$ and $\Delta \varphi$ windows for the 
cluster association.
The drawback is a significant decrease of the overall 
efficiency (see~\cite{PPR2} for details), which should be corrected on the 
basis of Monte Carlo simulations.

The magnetic field affects mostly the number of clusters in the second layer
and consequently the number of found tracklets decreases with 
increasing field strength.
This is mainly due to the tracks of very low momentum which are unable to hit
the second layer of SPD.
In principle, a special running session with the magnetic field off will 
offer the best configuration for the multiplicity measurement.

A more refined multiplicity measurement in the central barrel could 
be obtained with track reconstruction using all the six layers of the ITS,
the Time Projection Chamber (TPC) and the Transition Radiation Detector (TRD)
at the price of a smaller pseudorapidity coverage (($|\eta|<0.9$) and a longer
analysis time due to stricter requirements on alignment and detector 
calibration. 

In the forward and backward regions, the multiplicity will be measured by
the Forward Multiplicity Detectors (FMD) which are 5 silicon strip ring
counters with 51200 channels covering the regions $-5.1<\eta<-1.7$ and 
$1.7<\eta<3.4$.
Due to the high occupancy expected for central Pb-Pb interactions 
(up to 2.2 particles per pad for \dndemax=6000) it will not be possible to 
determine the number of charged particles by simply counting the pads which 
have fired, due to the large contribution of channels with 
more than one incident particle.
Charged multiplicity will be therefore estimated from the measurement
of the total energy deposited in each pad.
An alternative method based on counting empty pads and extracting from it 
the number of incident particles assuming Poisson statistics has also been 
developed.

\begin{figure}[bt]
\centering
\resizebox{\textwidth}{!}{
\includegraphics*[]{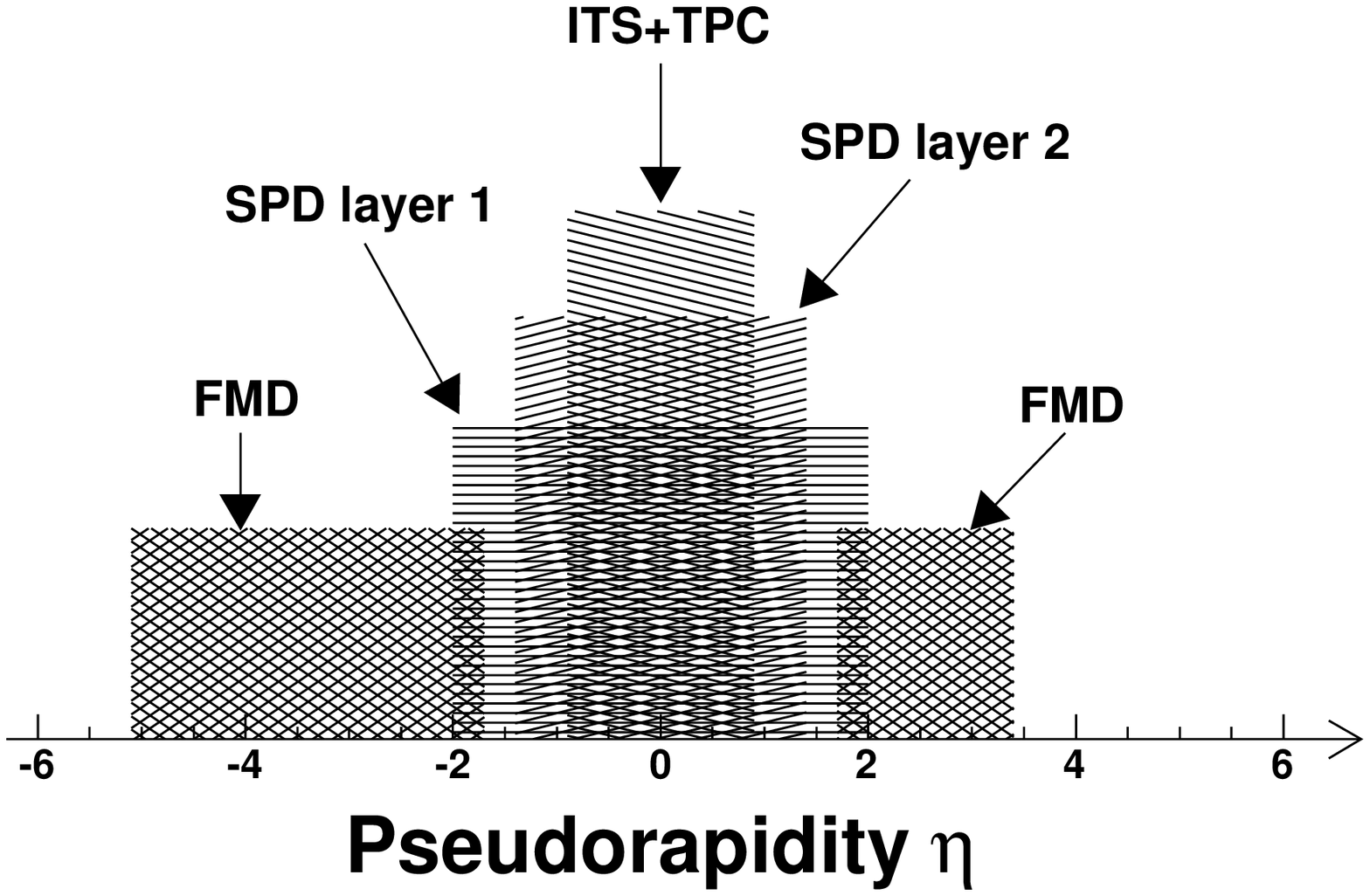}
\includegraphics*[]{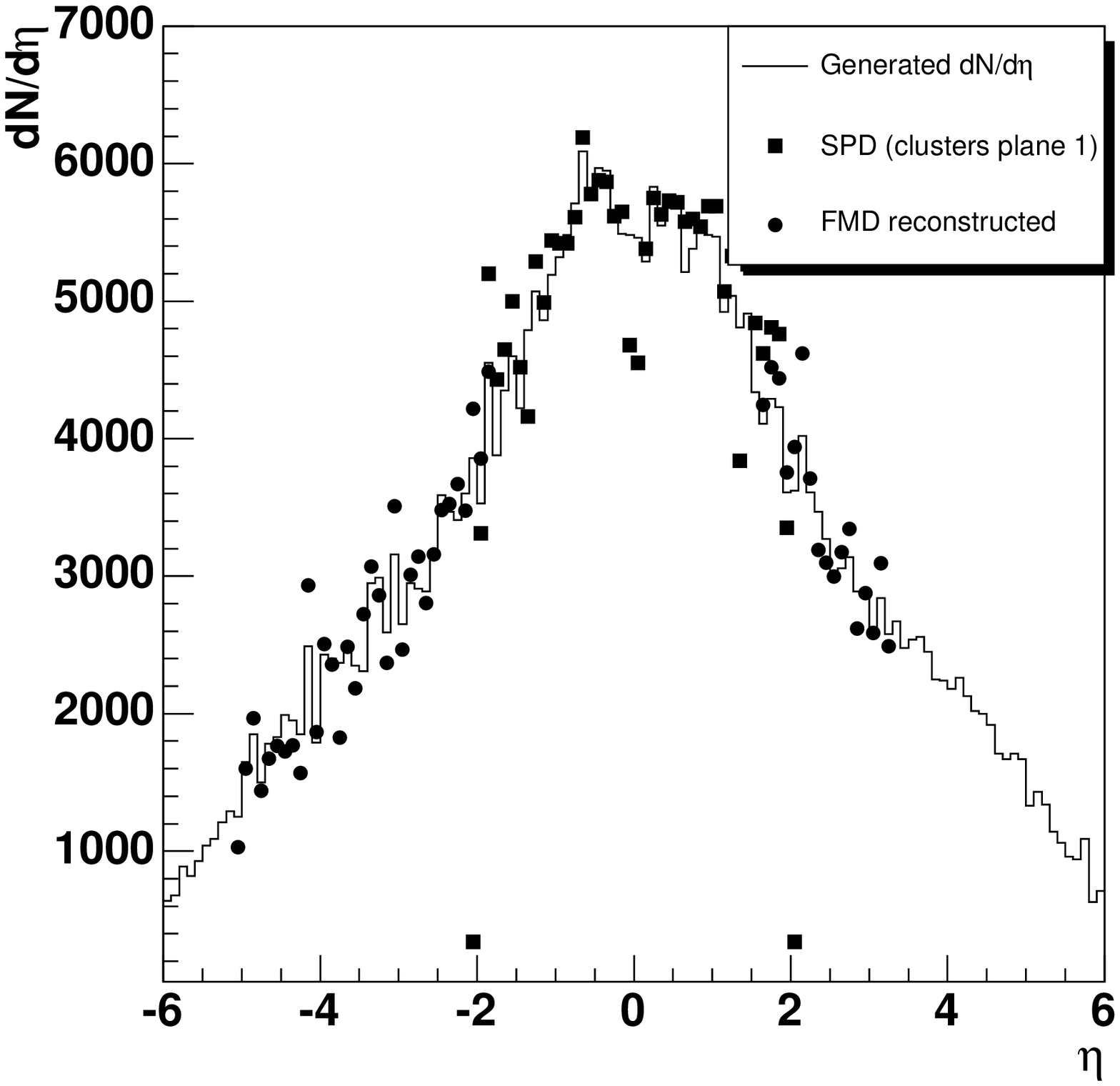}
}
\caption{Left panel: pseudorapidity coverage of the ALICE detectors 
used in multiplicity measurements. 
Right panel: $\eta$ distribution of charged particles reconstructed in the 
ITS and FMD detectors for a central HIJING event.}
\label{fig:etacov}
\end{figure}

The ITS and FMD detectors will allow to measure charged particles over a wide 
(about 8 $\eta$-units) pseudorapidity region, as it can be seen in 
fig.~\ref{fig:etacov} (left).
It should be noted that the acceptances quoted in this plot are referred to
the nominal vertex position and that the spread of the vertex position along
the beam direction will allow to extend the $\eta$ coverage.
In fig.~\ref{fig:etacov} (right) the generated and reconstructed \dnde\
pseudorapidity distributions of charged particles for a single central 
HIJING event are shown.
One can see that even for a single event the accuracy on the multiplicity 
determination, when using a bin width of 0.1 $\eta$-units, is of the
order of 7\%.
The few bins in the ITS acceptance regions where one observes a clear 
underestimation of the multiplicity are simply due to the geometrical junctions
of the modules of the SPD.

\section*{Acknowledgments}
I wish to thank the NA50 and ALICE collaborations.
I'm also grateful to Gunther Roland (PHOBOS) and Tiziano Virgili (NA57)
for giving me plots and material for the presentation.
Thanks also to M. Idzik, M. Monteno, M. Nardi and L. Ramello for fruitful
discussions and advices.

\end{document}